\documentclass[%
 reprint,
superscriptaddress,
 amsmath,amssymb,
 aps,
]{revtex4-2}

\usepackage{graphicx}
\usepackage{dcolumn}
\usepackage{bm}
\usepackage{hyperref}

\usepackage{lipsum}

\DeclareMathAlphabet{\mathpzc}{OT1}{pzc}{m}{it}
 
\def\beq{\begin{equation}}
\def\eeq{\end{equation}}
\def\bea{\begin{eqnarray}}
\def\eea{\end{eqnarray}}
\def\bwt{\begin{widetext}}
\def\ewt{\end{widetext}}
\def\nin{\noindent} 
\def\nn{\nonumber\\}

\begin{document}


\title{Investigating the imprint of quintessence in cosmic magnification}



\author{Enas Mohamed}\email{enasibrahim258@gmail.com}
\affiliation{Department of Physics \& Astronomy, Botswana International University of Science and Technology, Palapye, Botswana}

\author{Didam G.A. Duniya}\email{duniyaa@biust.ac.bw}
\affiliation{Department of Physics \& Astronomy, Botswana International University of Science and Technology, Palapye, Botswana}

\author{Hassan Abdalla}\email{hassanahh@gmail.com}
\affiliation{1IPARCOS and Department of EMFTEL, Universidad Complutense de Madrid, E-28040 Madrid, Spain}
\affiliation{Centre for Space Research, North-West University, Potchefstroom 2520, South Africa}
\affiliation{Department of Astronomy and Meteorology, Omdurman Islamic University, Omdurman 382, Sudan}

\author{Bishop Mongwane}\email{bishop.mongwane@uct.ac.za}
\affiliation{Department of Mathematics \& Applied Mathematics, University of Cape Town, South Africa}

\date{\today}

\begin{abstract}\nin
We study cosmic magnification beyond lensing in a late-time universe dominated by quintessence and cold dark matter. The cosmic magnification angular power spectrum, especially going beyond the well-known lensing effect, provides an independent avenue for investigating the properties of quintessence, and hence, dark energy. By analysing the magnification power spectrum at different redshifts, it is possible to extract new information about the large-scale imprint of dark energy, including whether we are able to disentangle different models from one another. Using three well-known quintessence models, we analyse the cosmic magnification angular power spectrum while taking relativistic corrections into account. We found that it will be difficult to distinguish between quintessence models, and quintessence from the cosmological constant, in lensing magnification angular power spectrum on large scales, at redshifts $z \,{\leq}\, 1$; whereas, when relativistic corrections are incorporated, the total magnification angular power spectrum holds the potential to distinguish between the models, at the given $z$. At $z \,{\geq}\, 3$, the lensing magnification angular power spectrum can be a reasonable approximation of the total magnification angular power spectrum. We also found that both the total relativistic and the Doppler magnification signals, respectively, surpass cosmic variance at $z \,{\leq}\, 0.5$: hence the effect may be detectable at the given $z$. On the other hand, the ISW and the time-delay magnification signals, respectively, are surpassed by cosmic variance on all scales, at epochs up to $z \,{=}\, 4.5$, with the gravitational-potential magnification signal being zero. 
\end{abstract}


\maketitle

\section{Introduction}\label{sec:intro}

The discovery of the expansion of the Universe \cite{Lundmark:1925mn, Lemaitre:1927zz, Hubble:1929ig, Steer:2012qs} is one of the most fundamental and profound findings in modern cosmology. The evidence of this expansion has been established observationally \cite{Lundmark:1925mn} and theoretically \cite{Lemaitre:1927zz}, with a further observational proof \cite{Hubble:1929ig}. Recently, it was further discovered that this expansion is accelerating \cite{SupernovaSearchTeam:1998fmf, SupernovaCosmologyProject:1998vns} at late cosmic times. Since the discovery of the cosmic acceleration, which is primarily attributed to dark energy (e.g. \cite{SupernovaSearchTeam:1998fmf, SupernovaCosmologyProject:1998vns, Copeland:2006wr, Bamba:2012cp, Tsujikawa2011, Tsujikawa:2013fta, Duniya:2013eta, Ratra:1987rm, Brax:1999gp, Barreiro:1999zs, Duniya:2015nva}), understanding its origin and nature on a fundamental level has become a central pursuit in cosmological research. While a cosmological constant $\Lambda$ (e.g. \cite{SupernovaSearchTeam:1998fmf, SupernovaCosmologyProject:1998vns, Copeland:2006wr, Bamba:2012cp}) has been a simple and compelling explanation for dark energy, recent theoretical and observational advances have challenged this simplistic, static paradigm and sparked a growing interest in more dynamic forms of dark energy, such as quintessence (e.g. \cite{Copeland:2006wr, Tsujikawa2011, Bamba:2012cp, Tsujikawa:2013fta, Duniya:2013eta, Ratra:1987rm, Brax:1999gp, Barreiro:1999zs}). Quintessence posits that dark energy is a scalar field $\varphi$ evolving on a potential $U(\varphi)$ with a time-varying equation of state parameter $w_\varphi$, differing from that of the cosmological constant ($w_\Lambda \,{=}\, {-}1$), which is characterized by a fixed energy density $\rho_\Lambda \,{\propto}\, \Lambda$. The quintessence model is a self-consistent model, with well-defined cosmological parameters: $w_\varphi$, which governs the background cosmology, and $c_{s\varphi}$, the physical sound speed, which governs the perturbation dynamics. These parameters are specified ad hoc for fluid or phenomenological dark energy. 

One of the challenging aspects of probing quintessence is disentangling its properties from those of other theories in cosmological observables, such as the galaxy power spectrum. However, cosmic magnification \cite{Zhan:2017uwu, Bonvin:2008ni, Schneider:2006eta, Refregier:2003ct, Schneider:2006gl, Duniya:2016ibg, Duniya:2016gcf, Duniya:2022vdi, Duniya:2023xgx, Bacon:2014uja, Raccanelli:2016avd, Bonvin:2016dze, Chen:2018hil, Andrianomena:2018aad, Coates:2020jzw, Duniya:2022xcz, Jeong:2011as, Hildebrandt:2015kcb, Blain:2001yf, Ziour:2008awn, Schmidt:2009rh, Schmidt:2010ex, Liu:2013yna, Camera:2013fva, Bartelmann:1999yn, Hui:2007cu, Hui:2007tm, Bonvin:2008ni, VanWaerbeke:2009fb, Duncan:2013haa, Gillis:2015caa, Umetsu:2015baa} can provide an independent, complementary avenue to probe the imprint of quintessence. Cosmic magnification is sourced by different cosmological effects in the large-scale structure. The commonly known and most studied source of cosmic magnification encapsulates cosmic magnification as an effect arising from the bending of light by gravitational fields of intervening or foreground objects: a phenomenon known as (gravitational) lensing \cite{Zhan:2017uwu, Schneider:2006eta, Bonvin:2008ni, Refregier:2003ct, Schneider:2006gl}. This effect has become a powerful tool in modern cosmology, enabling the study of the distribution of dark matter and the geometry of the Universe on large scales. While typically used to probe the large-scale structure, recent work has demonstrated that lensing magnification can also provide insights into the properties of dark energy \cite{Zhan:2017uwu}. Holistically, however, cosmic magnification extends beyond lensing magnification. It includes \cite{Duniya:2016ibg, Duniya:2016gcf, Duniya:2022vdi, Duniya:2023xgx}: Doppler effect (e.g. \cite{Bonvin:2008ni, Bacon:2014uja, Raccanelli:2016avd, Bonvin:2016dze, Chen:2018hil, Andrianomena:2018aad, Coates:2020jzw}), integrated-Sachs-Wolfe (ISW) effect (e.g. \cite{LoVerde:2006cj, Raccanelli_2012, Ballardini:2018cho}), time-delay effect (e.g. \cite{Raccanelli:2013gja, Baklanov:2020edg, Liao:2020yqz, Chan:2021jhh, Bayer:2021ugw, Baldwin:2021lud, Er:2022lad}), and gravitational-potential effect. These contributions need to be incorporated in the analysis in order to realise the full potential of cosmic magnification as a cosmological probe.

The potential for cosmic magnification to serve as a probe for quintessence lies in its sensitivity to the background expansion rate of the Universe and its dependence on the distribution of mass or energy and peculiar velocities. The interplay between these quantities, via the Raychaudhuri equation, the Poisson equation, and the Euler equation, connects cosmic magnification and quintessence. In essence, quintessence, by influencing the background expansion, modifies the formation and evolution of cosmic structures, thereby affecting the total magnification signal. Specifically, $w_\varphi$ affects the time-rate of growth of cosmic structures, such as galaxy clusters, which in turn alters the magnification of background sources via the gravitational potential (lensing, ISW, time-delay, and gravitational-potential effects, respectively) and velocity potential (Doppler effect). Probing quintessence with cosmic magnification, especially going beyond lensing, presents an exciting new avenue to investigate the properties of dark energy, in both background and perturbation. By studying how the magnification signals vary at different redshifts, it is possible to extract information about the large-scale imprint of quintessence, including whether we can disentangle different quintessence models from one another, and from other models.

In this paper we probe the imprint of quintessence in the cosmic magnification, beyond lensing---incorporating the Doppler, ISW, time-delay, and gravitational-potential corrections, respectively---using the angular power spectrum. The study seeks a theoretical understanding of the imprint of quintessence in cosmic magnification on large scales, using three well-studied quintessence models. Theoretically identifying the imprint of quintessence in the magnification angular power spectrum can help in correctly distinguishing it from other theories (including the $\Lambda$ model), and also in isolating bias in the extraction of parameters that pin the nature of dark energy, in the data. The rest of the paper is structured as follows. In \S\ref{sec:overdensity} we discuss the observed overdensity of (apparent) magnitude number counts, and in \S\ref{sec:Cls} we give the theoretical formalism of the magnification angular power spectrum. In \S\ref{sec:QCDM} we describe the Universe dominated by quintessence and cold dark matter: we discuss the background cosmology and give the numerical analysis of the magnification angular power spectrum, including comparison with the well-known cosmological constant model. We conclude in \S\ref{sec:Concl}.

\section{The Observed Magnification Overdensity}
\label{sec:overdensity}

Cosmic objects are observed in redshift space as projected structures or images on a transverse surface (relative to the line of sight), referred to as the screen or image plane, at a given redshift $z$ in a direction ${-\bf n}$. The apparent area of an image on the screen carries information of possible transformations undergone by the incident photons or their properties (e.g. flux). In a smooth, homogeneous universe cosmic objects will be observed at their true position, with a one-to-one mapping between the apparent image plane and the true image plane (at the object), and the transverse area remains unchanged. 

However, in an inhomogeneous universe the screen area is distorted by a factor $\mu$, such that \cite{Duniya:2016ibg, Duniya:2016gcf, Duniya:2022vdi}
\bea\label{mu}
\mu^{-1}({\bf n},z) = \dfrac{\tilde{\cal A}({\bf n},z)}{\bar{\cal A}(\bar{z})} , 
\eea 
where $\mu$ gives the magnification coefficient, $\tilde{\cal A} \,{=}\, \partial{A}/\partial\Omega_{\bf n}$ is the apparent or observed area per unit solid angle (in redshift space) and $\bar{\cal A}$ is the background term. The observational implications of \eqref{mu} are that (i) overdensed regions will have a magnification factor $\mu \,{>}\, 1$ and objects will appear closer than they truly are, with the apparent image-plane area appearing to be reduced, (ii) underdensed regions will have $\mu \,{<}\, 1$ and objects will appear farther away, with the apparent screen area appearing to be enlarged, and (iii) even regions will have $\mu \,{=}\, 1$ and objects are seen at their true position, with the apparent screen area remaining unchanged. A consequence of \eqref{mu} is that the background flux is transformed:
\beq\label{flux}
\bar{\cal F}(\bar{z}) \to \mu({\bf n},z) \bar{\cal F}(\bar{z}) = \tilde{\cal F}({\bf n},z),
\eeq 
where $\tilde{\cal F}$ is the observed flux per unit solid angle per redshift, with  $\bar{\cal F}$ being the background (unperturbed) term. 

Thus, by observing the distribution of magnified sources, i.e. the magnitude number counts, observers can obtain the observed (relativistic) magnification overdensity (e.g. \cite{Duniya:2016ibg, Duniya:2016gcf, Duniya:2022xcz, Duniya:2022vdi, Duniya:2023xgx, Jeong:2011as, Hildebrandt:2015kcb}), given by
\bea\label{Delta:obs} 
\Delta^{\rm obs}_{\cal M}({\bf n},z)  = {\cal Q}(z)\, \Big[\mu({\bf n},z) - 1\Big],
\eea 
where ${\cal Q}$ is the \emph{magnification bias} \cite{Duniya:2016ibg, Blain:2001yf, Ziour:2008awn, Schmidt:2009rh, Schmidt:2010ex, Liu:2013yna, Camera:2013fva, Hildebrandt:2015kcb}, given by
\beq\label{magbias}
{\cal Q} \equiv \left. \dfrac{\partial\ln\bar{\cal N}}{\partial\ln\bar{\cal F}} \right|_z = \dfrac{1}{2}\left(2-5\alpha\right),
\eeq
where $\bar{\cal N} = \bar{n}\bar{\cal F}$ is the magnitude number-count density in the background, $\bar{n}$ is the differential magnitude number counts, i.e. number per unit flux, $\bar{\cal F}$ is as in \eqref{flux}, and $\alpha$ is the logarithmic slope of the differential magnitude number counts, given by
\beq\label{dlogndm}
\alpha(m) = \dfrac{\partial}{\partial{m}} \log_{10}\bar{n}(m),
\eeq
with $m$ being the apparent magnitude:
\beq\nonumber
m = m_* - 2.5\log_{10}\left(\dfrac{\bar{\cal F}}{\bar{\cal F}_*}\right),
\eeq
where $m_*$ and $\bar{\cal F}_*$ are fixed reference values (constants).

In order to compute the magnification coefficient $\mu$, we need to connect the redshift-space area density $\tilde{\cal A}$ to the real-space area density ${\cal A}$ in the physical area element:
\beq\label{dA1}
dA = {\cal A} (\theta_O,\vartheta_O)\, d\theta_O d\vartheta_O, 
\eeq
where $\theta_O$ and $\vartheta_O$ are the zenith and the azimuthal angles, respectively, at the observer $O$.  General Relativity allows coordinate freedom. However, the change of coordinates causes the perturbations to also change. How then do we deal with this coordinate freedom? One way is to fix the spacetime metric by choosing an appropriate constraint or gauge. Another way is to assume a general spacetime metric, and then define gauge-invariant perturbation quantities. In this work, we adopt the latter approach.

The area density ${\cal A}$ lives in a spacetime metric:
\beq\label{ds2}
ds^2 = g_{\beta\nu}dx^\beta dx^\nu ,
\eeq
which is prescribed by the tensor $g_{\beta\nu}$ (with $\beta$ and $\nu$ being spacetime indices), given by 
\begin{align}\label{g_00}
g_{00}(\eta,x^i) =& -a^2 [1+2\phi(\eta,x^i)],\\ \label{g_0j}
g_{0j}(\eta,x^i) =&\, a^2 \partial_j B(\eta,x^i) = g_{j0}(\eta,x^i), \\ \label{g_ij}
g_{ij}(\eta,x^i) =&\, a^2\left[1-2D(\eta,x^i)\delta_{ij} + 2{\cal D}_{ij}E(\eta,x^i)\right],
\end{align}
where $a \,{=}\, a(\eta)$ is the spacetime scale factor, $\eta$ is conformal time, $x^i$ denotes the real spatial coordinates and, $\phi$, $B$, $D$, and $E$ are the scalar perturbative degrees of freedom in $g_{\beta\nu}$; with ${\cal D}_{ij} \,{=}\, \partial_i\partial_j \,{-}\, \frac{1}{3}\delta_{ij}\nabla^2$. (Other notations retain their standard definitions.) Note that \eqref{g_00}--\eqref{g_ij} describe a general metric, with exhausted scalar degrees of freedom---which are coordinate-dependent. 

 Consider an arbitrary coordinate change from an initial set of coordinates $x^\beta \,{=}\, (\eta,\, x^i)$ to a new set of coordinates $\hat{x}^\beta \,{=}\, (\hat{\eta},\, \hat{x}^i)$, given by
\beq\label{coordchange-1}
x^\beta \to \hat{x}^\beta = x^\beta + \xi^\beta(x^\nu),
\eeq
where $\xi^\beta \,{=}\, (\xi^0,\, \xi^i)$ is the (perturbation) transformation 4-vector, with $|\xi^\beta| \,{\ll}\, 1$, and
\beq\nonumber
\hat{\eta} = \eta + \xi^0(\eta,x^j),\quad \hat{x}^i = x^i + \xi^i(\eta,x^j),
\eeq
with $\xi^0$ measuring the ``time shift'' between adjacent constant-$\eta$ hypersurfaces and, $\xi^i \,{=}\, \partial^i \xi$ giving the ``space shift'' between spatial coordinates on the hypersurface. (The 3-vector $\xi^i$ is taken as a scalar-generated vector, to correspond to the perturbations in $g_{\beta\nu}$, with $\xi$ being a scalar.) Consequently, the coordinate transformation \eqref{coordchange-1} leads to the ``gauge transformation law'' for the metric-tensor perturbations, given by
\beq\label{gtranslaw}
\delta{g}_{\beta\nu}(\eta,x^i) \to \delta\hat{g}_{\beta\nu}(\eta,x^i) = \delta{g}_{\beta\nu}(\eta,x^i) - L_\xi\bar{g}_{\beta\nu}(\bar{\eta}),
\eeq
where henceforth, $\delta{X}$ denotes first-order perturbation in parameter $X$, and $L_\xi$ is the Lie derivative with respect to $\xi$:
\beq\label{Liederiv}
L_\xi\bar{g}_{\beta\nu} = \xi^\lambda\partial_\lambda\bar{g}_{\beta\nu} + \bar{g}_{\lambda\nu}\partial_\beta\xi^\lambda + \bar{g}_{\beta\lambda}\partial_\nu\xi^\lambda,
\eeq
so that ($\lambda$ being a 4-index) the metric tensor components \eqref{g_00}--\eqref{g_ij} transform to retain the same form, by
\begin{align}\label{hat_g_00}
\hat{g}_{00}(\eta,x^i) =& -a^2 [1+2\hat{\phi}(\eta,x^i)],\\ \label{hat_g_0j}
\hat{g}_{0j}(\eta,x^i) =&\, a^2 \partial_j \hat{B}(\eta,x^i) = \hat{g}_{j0}(\eta,x^i), \\ \label{hat_g_ij}
\hat{g}_{ij}(\eta,x^i) =&\, a^2\left[1-2\hat{D}(\eta,x^i)\delta_{ij} + 2{\cal D}_{ij} \hat{E}(\eta,x^i)\right],
\end{align}
where $\hat{\phi}$, $\hat{B}$, $\hat{D}$, and $\hat{E}$ are the gauge-transforms of $\phi$, $B$, $D$, and $E$, respectively, given by
\bea\label{hat-phi}
\hat{\phi} &=& \phi - {\cal H}\xi^0 - \xi^0{'},\\ \label{hat-B}
\hat{B} &=& B + \xi^0 - \xi',\\ \label{hat-D}
\hat{D} &=& D +\dfrac{1}{3}\nabla^2\xi + {\cal H}\xi^0,\\ \label{hat-E}
\hat{E} &=& E - \xi,
\eea
with a prime denoting derivative with respect to $\eta$, and ${\cal H} \,{=}\, a'/a$. Similarly, \eqref{coordchange-1} leads to a gauge transformation law for the 4-velocity (and energy-momentum tensor), so that
\beq\label{hat-du}
\delta{u}^\beta \to \delta\hat{u}^\beta = \delta{u}^\beta - L_\xi \bar{u}^\beta = a^{-1}\left(-\hat{\phi}, \partial^i \hat{v}\right),
\eeq
where $\hat{\phi}$ is as in \eqref{hat-phi} (see Appendix~\ref{App:Inv}), and
\beq\label{hat-v}
\hat{v} = v + \xi' ,
\eeq
with $v$ being the coordinate velocity potential. Thus, the 4-velocity of a fundamental observer living in the metric \eqref{hat_g_00}--\eqref{hat_g_ij}, is given by
\beq\label{hat-u}
\hat{u}^\beta = a^{-1}\left(1 -\hat{\phi}, \partial^i \hat{v}\right),
\eeq
where the background term $\bar{u}^\beta \,{=}\, a^{-1}\delta^\beta\-_0$.

The gauge transformation law for the area density perturbations (being scalars), is given by
\beq\label{dA-trans}
\delta\tilde{\cal A} = \delta{\cal A} - \xi^\beta \partial_\beta \bar{\cal A} = \delta{\cal A} - \delta\eta \bar{\cal A}',
\eeq
where $\delta\tilde{\cal A}$ and $\delta{\cal A}$ are the redshift-space and real-space area density perturbations, respectively (with $\xi^0$ giving the perturbation in $\eta$). At first order, we have $\delta\eta \,{=}\, (\partial\bar{\eta}/\partial\bar{z}) \delta{z}$, and the magnifcation coefficient \eqref{mu}, becomes
\beq\label{Mag2}
\mu^{-1} = 1 + \dfrac{\delta{\cal A}}{\bar{\cal A}} + 2\left(1 - \dfrac{1}{\bar{r} {\cal H}}\right) \dfrac{\delta{z}}{1+\bar{z}}, 
\eeq
where
\begin{equation}\label{eq:A/z}
\frac{\partial \bar{\mathcal{A}}}{\partial \bar{z}} = 2a \left(1- \frac{1}{\mathcal{H}\bar{r}}\right) \bar{\mathcal{A}},
\end{equation} 
with $\bar{r}$ being the background radial comoving distance, and $\delta{z} \,{=}\, z \,{-}\, \bar{z}$ is the perturbation in the observed redshift,  and $\bar{\mathcal{A}}$ is as given in (\ref{delta-A}).

 As previously stated, the change of coordinates \eqref{coordchange-1} transforms the cosmological perturbations as in \eqref{hat-phi}--\eqref{hat-E} and \eqref{hat-v}, which change with the choice of $\xi^\beta$. To deal with this arbitrariness, we define gauge-independent quantities. Hence by imposing the condition
\beq\label{gauge2}
\hat{B} = \hat{E} = 0,
\eeq 
and using \eqref{hat-phi}--\eqref{hat-E} and \eqref{hat-v}, we therefore define:
\bea\label{inv-Phi}
\Phi &\equiv & \phi - {\cal H}\left(E'-B\right) + B' - E'', \\ \label{inv-Psi}
\Psi &\equiv & D + \dfrac{1}{3}\nabla^2 E + {\cal H}\left(E'-B\right), \\ \label{inv-V}
V &\equiv & v+E',
\eea
where $v$ is the gauge-dependent, coordinate velocity potential as in \eqref{hat-v}. The new metric potentials \eqref{inv-Phi} and \eqref{inv-Psi}, commonly known as the Bardeen potentials \cite{Bardeen:1980kt, Durrer:1993tti}, and the new velocity potential \eqref{inv-V} are gauge-invariant (see Appendix~\ref{App:Inv}). The associated 4-velocity of a fundamental observer, given \eqref{hat-u}, is 
	\beq\label{inv-U}
	U^\beta = a^{-1}\left(1-\Phi, \partial^i V\right),
	\eeq
	where $\Phi$ and $V$ are as given by \eqref{inv-Phi} and \eqref{inv-V}, respectively. Thus, \eqref{inv-Phi}--\eqref{inv-U} allow us to do cosmology in any spacetime coordinates and metric. 
	
Then after some calculations---taking the comoving radial distance as $r \,{=}\, \bar{r} \,{+}\, \delta{r}$, and the angles at a source $S$ as $\theta_S \,{=}\, \theta_O \,{+}\, \delta\theta$ and $\vartheta_S \,{=}\, \vartheta_O \,{+}\, \delta\vartheta$---we obtain the perturbations in ${\cal A}$ and $z$  (see e.g. \cite{Duniya:2016ibg, Duniya:2016gcf, Jeong:2011as}, for full details), given by
\bea\label{delta-A}
\dfrac{\delta{\cal A}}{\bar{\cal A}} &=& \int^{\bar{r}_S}_0 d\bar{r} \dfrac{\bar{r} - \bar{r}_S}{\bar{r}_S \bar{r}} \nabla^2_\Omega \Big(\phi + D + \dfrac{1}{3}\nabla^2 E + B' - E''\Big) \nn
&& +\; \dfrac{2}{\bar{r}_S} \int^{\bar{r}_S}_0 d\bar{r} \Big(\phi + D + \dfrac{1}{3}\nabla^2 E + B' - E''\Big) \nn
&& +\; \Big[\dfrac{2}{\bar{r}_S } \left(B-E'\right) -2D - \dfrac{2}{3}\nabla^2 E \Big]^S_O , \nn
&=& -2\Psi\Big|^S_O + \int^{\bar{r}_S}_0{ d\bar{r}\left[\dfrac{2}{\bar{r}_S} + \left(\bar{r} - \bar{r}_S\right) \dfrac{\bar{r}}{\bar{r}_S } \nabla^2_\perp \right] \left(\Phi + \Psi\right) } \nn
&& +\; 2{\cal H}\left(1 - \dfrac{1}{\bar{r}_S {\cal H}}\right) \Big[E'-B\Big]^S_O,
\eea
where $\bar{\cal A} \,{=}\, a^2\bar{r}^2\sin{\theta}_O$ is the background area density, $\bar{r}_S \,{=}\, \bar{r}(\bar{z}_S)$ being the background comoving distance at the source redshift $z_S$, and
	\beq\nonumber
	\dfrac{1}{\bar{r}^2} \nabla^2_\Omega = \nabla^2 - \dfrac{\partial^2}{\partial\bar{r}^2} - \dfrac{2}{\bar{r}} \dfrac{\partial}{\partial\bar{r}} \;\equiv\; \nabla^2_\perp,
	\eeq
	with $\nabla^2$ being the standard Laplacian; $\Phi$ and $\Psi$ are as in \eqref{inv-Phi} and \eqref{inv-Psi}, and we have (see e.g. \cite{Duniya:2016ibg, Duniya:2016gcf, Jeong:2011as}, for full details)
\bea\label{delta-z}
\dfrac{\delta{z}}{1+\bar{z}} &=& - \int^0_{\bar{r}_S} d\bar{r} \Big[\phi' + D' + \dfrac{1}{3}\nabla^2 E' + (B' - E'')'\Big], \nn
&& +\; \Big[\phi + B' - E'' - \dfrac{\partial V}{\partial\bar{r}} \Big]^O_S , \nn
&=& -\left[\Phi + \Psi - \dfrac{\partial{V}}{\partial\bar{r}} - D - \dfrac{1}{3}\nabla^2 E \right]^S_O \nonumber\\
&& + \int^{\bar{r}_S}_0 d\bar{r} \left(\Phi' +\Psi'\right),
\eea
where $V$ is as in \eqref{inv-V}. (The rigorous derivations of \eqref{delta-A} and \eqref{delta-z}, are given by e.g. \cite{Duniya:2016ibg, Duniya:2016gcf, Jeong:2011as}.) Note that both the real-space area contrast $\delta{\cal A}/\bar{\cal A}$ and the redshift contrast $\delta{z}/(1 \,{+}\, \bar{z})$ are not directly observable individually, because of the last terms in the second line in \eqref{delta-A} and the last two terms in the square brackets in \eqref{delta-z}, respectively. However, their sum as given by \eqref{Mag2}, being the redshift-space area contrast $\delta\tilde{\cal A}/\bar{\cal A}$, is observable, which measures the amount of distortion in the screen area, in redshift space. (See Appendix \ref{App:Inv}.)

Thus, by combining \eqref{delta-A} and \eqref{delta-z} in  \eqref{Mag2}, the observed magnification overdensity \eqref{Delta:obs}, becomes
\begin{align} \label{Delta:obs2}
\Delta^{\rm obs}_{\cal M} =& -{\cal Q}\int^{\bar{r}_S }_0{ d\bar{r}\left[\dfrac{2}{\bar{r}_S} + \left(\bar{r} - \bar{r}_S\right) \dfrac{\bar{r}}{\bar{r}_S } \nabla^2_\perp \right] \left(\Phi + \Psi\right) } \nn
&+\; 2{\cal Q}\left(\dfrac{1}{{\cal H}\bar{r}_S} - 1\right) \left[\dfrac{\partial V}{\partial\bar{r}} - \int^{\bar{r}_S }_0{d\bar{r} \left(\Phi' + \Psi' \right)} \right] \nn
&+\; 2{\cal Q}\Psi + 2{\cal Q}\left(1 -\dfrac{1}{{\cal H}\bar{r}_S}\right) \Phi,
\end{align}
where we have dropped the limits on non-integral terms: hence now denoting relative values, i.e. the values at $S$ relative to those at $O$. The first line in \eqref{Delta:obs2} gives the lensing term, given by
\begin{align}\label{lensing} 
\Delta^{\rm lensing}_{\cal M} \equiv  \dfrac{(2-5\alpha)}{2} \int^{\bar{r}_S }_0{ d\bar{r}\left(\bar{r}_S - \bar{r} \right) \dfrac{\bar{r}}{\bar{r}_S} \nabla^2_\perp \left(\Phi + \Psi\right)},
\end{align}
where we used \eqref{magbias}, and the non-lensing terms, otherwise known as the ``relativistic'' corrections, are given by
\begin{align}\label{Doppler}
\Delta^{\rm Doppler}_{\cal M} \equiv\;& (2-5\alpha) \left(\dfrac{1}{{\cal H}\bar{r}_S} - 1 \right) V^\parallel_m , \\ \label{ISW}
\Delta^{\rm ISW}_{\cal M} \equiv\;& (2-5\alpha) \left( \dfrac{1}{{\cal H}\bar{r}_S}- 1 \right)\int^{\bar{r}_S}_0{d\bar{r}\, (\Phi'+\Psi')} , \\ \label{timedelay}
\Delta^{\rm timedelay}_{\cal M} \equiv\;& -\dfrac{(2-5\alpha)}{\bar{r}_S} \int^{\bar{r}_S}_0{ d\bar{r}\, (\Phi+\Psi) } , \\ \label{potential}
\Delta^{\rm potential}_{\cal M} \equiv\;& (2-5\alpha)\Psi + (2-5\alpha)\left(1 -\dfrac{1}{{\cal H}\bar{r}_S}\right) \Phi ,
\end{align}
where \eqref{Doppler} gives the Doppler correction, \eqref{ISW} gives the ISW correction, \eqref{timedelay} gives the time-delay correction, \eqref{potential} gives the gravitational (potential) correction, and $V^\parallel_m = \partial V_m/\partial{r}$ (assuming that galaxies trace the same path as the underlying matter). These corrections together \eqref{Doppler}--\eqref{potential}, are otherwise known as ``relativistic'' corrections.


\section{The Magnification Angular Power Spectrum}
\label{sec:Cls}
Cosmic magnification is measurable in the apparent flux, which becomes transformed \eqref{flux}, in redshift space. It is induced by two main factors: Firstly, the image-plane area is distorted in real space \eqref{delta-A} by lensing \eqref{lensing}, which deforms the observation angles, and by time delay \eqref{timedelay}, which distorts the radial comoving distance. Secondly, by observing on the past lightcone the observed redshift is distorted by Doppler effect \eqref{Doppler}, which introduces a relative peculiar velocity between the source and the observer. The observed redshift is also distorted by ISW effect \eqref{ISW}, which integrates the time-rate of change of the gravitational potential along the line of sight. Both the real-space area distortion and the redshift distortion are also partly induced by a gravitational-potential difference \eqref{potential}. These effects manifest in the magnification factor \eqref{Mag2}, in redshift space. Consequently, for (under) overdense regions we will have ($\mu \,{<}\, 1$) $\mu \,{>}\, 1$, resulting in the background flux being (demagnified) magnified. (See also discussion below \eqref{mu}.)

By using \eqref{Delta:obs2}--\eqref{potential}, the angular power spectrum of observed magnitude number counts (or flux distribution) at a source redshift $z \,{=}\, z_S$, is given by 
\beq\label{Cls}
C_\ell(z_S) = \dfrac{4}{\pi^2} \left(\dfrac{9}{10}\right)^2 \int{dk\, k^2 T(k)^2 P_{\Phi_p}(k) \Big|f_\ell(k,z_S) \Big|^2 },
\eeq
where (henceforth) we take that quintessence does not support anisotropic stress, and hence $\Psi \,{=}\, \Phi$, $P_{\Phi_p}$ is the power spectrum of the primordial gravitational potential $\Phi_p$, $k$ is the wavenumber, and
\begin{align}\label{f_ell}
f_\ell(k,z_S) &= (2-5\alpha) \int^{r_S }_0{ dr \dfrac{(r - r_S)}{r_S r} \ell(\ell+1) \check{\Phi}(k,r) j_\ell(kr)} \nn
& +\; (2-5\alpha) \left( \dfrac{1}{{\cal H}r_S } - 1\right) \check{V}^\parallel_m(k,z_S)  \dfrac{\partial j_\ell(kr_S)}{\partial(kr)}, \nn
& +\; 2(2-5\alpha) \left(1 - \dfrac{1}{{\cal H}r_S } \right) \int^{r_S}_0{dr \check{\Phi}'(k,r) j_\ell(kr)}  \nn
& +\; (2-5\alpha) \left(2 -\dfrac{1}{{\cal H}r_S }\right) \check{\Phi}(k,z_S) j_\ell(k,z_S) \nn
&-\; \dfrac{2(2-5\alpha)}{r_S} \int^{r_S }_0{ dr \check{\Phi}(k,r) j_\ell(kr)} ,
\end{align}
where we have dropped the bar from background terms, $\alpha$ is as in \eqref{dlogndm}, $j_\ell$ is the spherical Bessel function, with $\check{V}_m \,{=}\, V_m/\Phi_d$ (similarly for $\check{\Phi}$), $V_m$ is the gauge-invariant matter velocity potential, as in \eqref{inv-V}, $\Phi_d(k) \,{=}\, (9/10) \Phi_p(k) T(k)$ is the gravitational potential at the photon-matter decoupling epoch $z \,{=}\, z_d$, and $T(k)$ is linear transfer function \cite{Dodelson:2003bk}. The individual lines in \eqref{f_ell} give the contribution of lensing \eqref{lensing}, Doppler effect \eqref{Doppler}, ISW effect \eqref{ISW}, gravitational effect \eqref{potential}, and time-delay effect \eqref{timedelay}, respectively.


\section{The $\varphi$CDM Universe}
\label{sec:QCDM}

As previously stated, in this work we consider a $\varphi$CDM universe: dominated by quintessence $\varphi$ and cold dark matter (CDM). (See Appendix~\ref{QCDM:Eqns}, and references therein, for the cosmological equations.) The key parameters that need to be specified are the equation of state parameter $w_\varphi$ and physical sound speed $c_{s\varphi}$. These parameters are specified ad hoc for phenomenological dark energy. However, quintessence is self-consistent in that for any potential $U(\varphi)$ the parameters are determined by
\beq\label{w_q}
w_\varphi = \dfrac{\varphi'^2-2a^2U} {\varphi'^2+2a^2U},\quad c_{s\varphi} = 1,
\eeq
where $a$ is as in \S\ref{sec:overdensity}, and $\varphi$ evolves by the Klein-Gordon equation, given by
\beq\label{d2Vardt2}
\varphi'' + 2{\cal H}\varphi' +a^2\frac{\partial U}{\partial\varphi} = 0,
\eeq
with ${\cal H}$ and a prime being as in \S\ref{sec:overdensity}.

\begin{figure*}\centering
\includegraphics[scale=0.45]{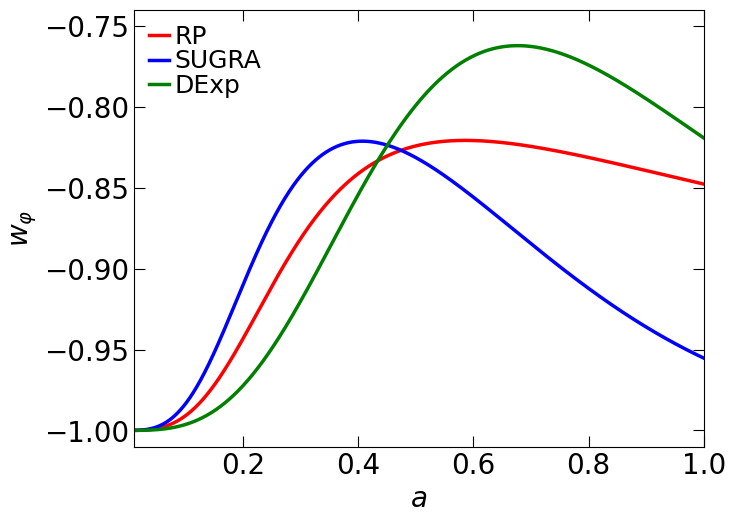} 
\caption{The plots of the equation of state parameters \eqref{w_RP}, \eqref{w_SUGRA}, and \eqref{w_DExp} for RP, SUGRA, and DExp models, respectively, with respect to scale factor $a$. These govern the background evolution of quintessence in the respective models.}\label{fig:w_x}
\end{figure*}

\subsection{The Quintessence Models}\label{subsec:QCDM}
The different quintessence models are specified by unique potentials $U$. Here we consider the most studied three models in the literature.

\subsubsection*{Ratra-Peebles model}
The Ratra-Peebles (RP) inverse potential, given by (e.g. \cite{Copeland:2006wr, Duniya:2013eta, Ratra:1987rm})
\beq\label{RP}
U(\varphi) = \dfrac{M^{4+\sigma}}{\varphi^{\sigma}},\quad \sigma > 0,
\eeq
where $M$ is a mass scale, and $\sigma$ is a free constant. Thus, we obtain
\beq\label{w_RP}
w_\varphi = \dfrac{y^2-6a^2A_0/x^\sigma} {y^2+6a^2A_0/x^\sigma},\quad A_0 \equiv \dfrac{\kappa^{2+\sigma}M^{4+\sigma}}{3H^2_0},
\eeq
where $A_0$ is a dimensionless constant, and
\beq\label{xy} 
x \equiv \kappa \varphi,\qquad y \equiv \kappa \varphi' /H_0, \qquad \kappa \equiv \sqrt{8\pi G},
\eeq
with $x$ and $y$ being the dimensionless quintessence field and its time derivative, respectively; $G$ being the Newton's gravitational constant. The Klein-Gordon equation \eqref{d2Vardt2}, for the RP potential, becomes
\beq\label{KG-RP}
y' + 2{\cal H}y = 3\sigma A_0 H_0 \dfrac{a^2}{x^{\sigma+1}} .
\eeq
For all numerical computations we set $A_0 \,{=}\, 0.58$ \cite{Duniya:2013eta} and $\sigma \simeq 0.52604704$.

\subsubsection*{Super-gravity model}
The super-gravity (SUGRA) potential (e.g. \cite{Copeland:2006wr, Duniya:2013eta, Brax:1999gp}), being a super-gravity correction \cite{Brax:1999gp} to the RP potential \eqref{RP}, is given by
\begin{align}\label{SUGRA}
U(\varphi) =\; \dfrac{M^{4+\sigma}}{\varphi^{\sigma}}\, e^{\frac{1}{2}\kappa^2 \varphi^2},
\end{align}%
where all parameters are as in \eqref{SUGRA}. Similarly, we obtain
\beq\label{w_SUGRA}
w_\varphi = \dfrac{y^2-6a^2A_0 e^{\frac{1}{2}x^2}/x^\sigma} {y^2+6a^2A_0 e^{\frac{1}{2}x^2}/x^\sigma}, 
\eeq
where $A_0$, $x$ and $y$ are as in \eqref{w_RP} and \eqref{xy}. The Klein-Gordon equation \eqref{d2Vardt2}, for the SUGRA potential, becomes
\beq\label{KG-SUGRA}
y' + 2{\cal H}y = 3 A_0 H_0 \left(\sigma-x^2\right) \dfrac{a^2 e^{\frac{1}{2}x^2}}{x^{\sigma+1}} ,
\eeq
and for all numerical computations, we chose $A_0 \,{=}\, 0.45$ \cite{Duniya:2013eta} and $\sigma \simeq 0.65451771$.

\subsubsection*{Double exponential model}
The double exponential (DExp) potential, is given by (e.g. \cite{Duniya:2013eta, Barreiro:1999zs})
\beq\label{DExp}
U(\varphi) = M^4_1\, e^{\sigma\kappa\varphi} + M^4_2 \, e^{\lambda\kappa\varphi} ,
\eeq%
where $M_1$ and $M_2$ are mass scales, and $\sigma$ and $\lambda$ are free constants, with $\kappa$ as in \eqref{xy}. Thus, we obtain 
\beq\label{w_DExp}
w_\varphi = \dfrac{y^2-6a^2A_0 \left(e^{\sigma x}+\gamma e^{\lambda x}\right) } {y^2+6a^2A_0 \left(e^{\sigma x}+\gamma e^{\lambda x}\right)}, \quad A_0 \equiv \dfrac{\kappa^2 M^4_1}{3H^2_0},
\eeq
where $\gamma \,{\equiv}\, M_2/M_1$ is a dimensionless constant, and the Klein-Gordon equation \eqref{d2Vardt2}, for the DExp potential, is
\beq\label{KG-DExp}
y' + 2{\cal H}y = -3 A_0 H_0 a^2 \left(\sigma e^{\sigma x}+\gamma\lambda e^{\lambda x}\right) ,
\eeq
with $x$ and $y$ being as in \eqref{xy}. Here we use $A_0 \,{=}\, 1.27$, $\sigma \simeq -4.09$, $\gamma \,{=}\, 0.27843809$, and $\lambda \,{=}\, 0.01$ for all numerical computations.

\subsection{The Background Evolutions}
By using \eqref{RP}--\eqref{KG-DExp}, the cosmological equations (see Appendix~\ref{QCDM:Eqns}) can be solved for the RP model given \eqref{w_RP} and \eqref{KG-RP}, for the SUGRA model given \eqref{w_SUGRA} and \eqref{KG-SUGRA}, and for the DExp model given \eqref{w_DExp} and \eqref{KG-DExp}. The sound speed \eqref{w_q} remains unity regardless of the choice of the quintessence potential. The goal here is to probe the chosen quintessence models on large scales. A suitable approach is to normalise the models at today ($z \,{=}\, 0$), i.e. to set the models to the same background universe at today. The consequence of this is that any characteristic signatures owing to the perturbations in the different models will be isolated on the large scales, in the magnification angular power spectrum. Hence we choose the quintessence parameters such that we have the same values $\Omega_{m0} \,{=}\, 0.3$ and $H_0 \,{=}\, 67.8$ km/s/Mpc \cite{Planck:2018vyg}, for the three quintessence models. We evolve the cosmological equations from the photon-matter decoupling epoch.

\begin{figure*}\centering
\includegraphics[scale=0.4]{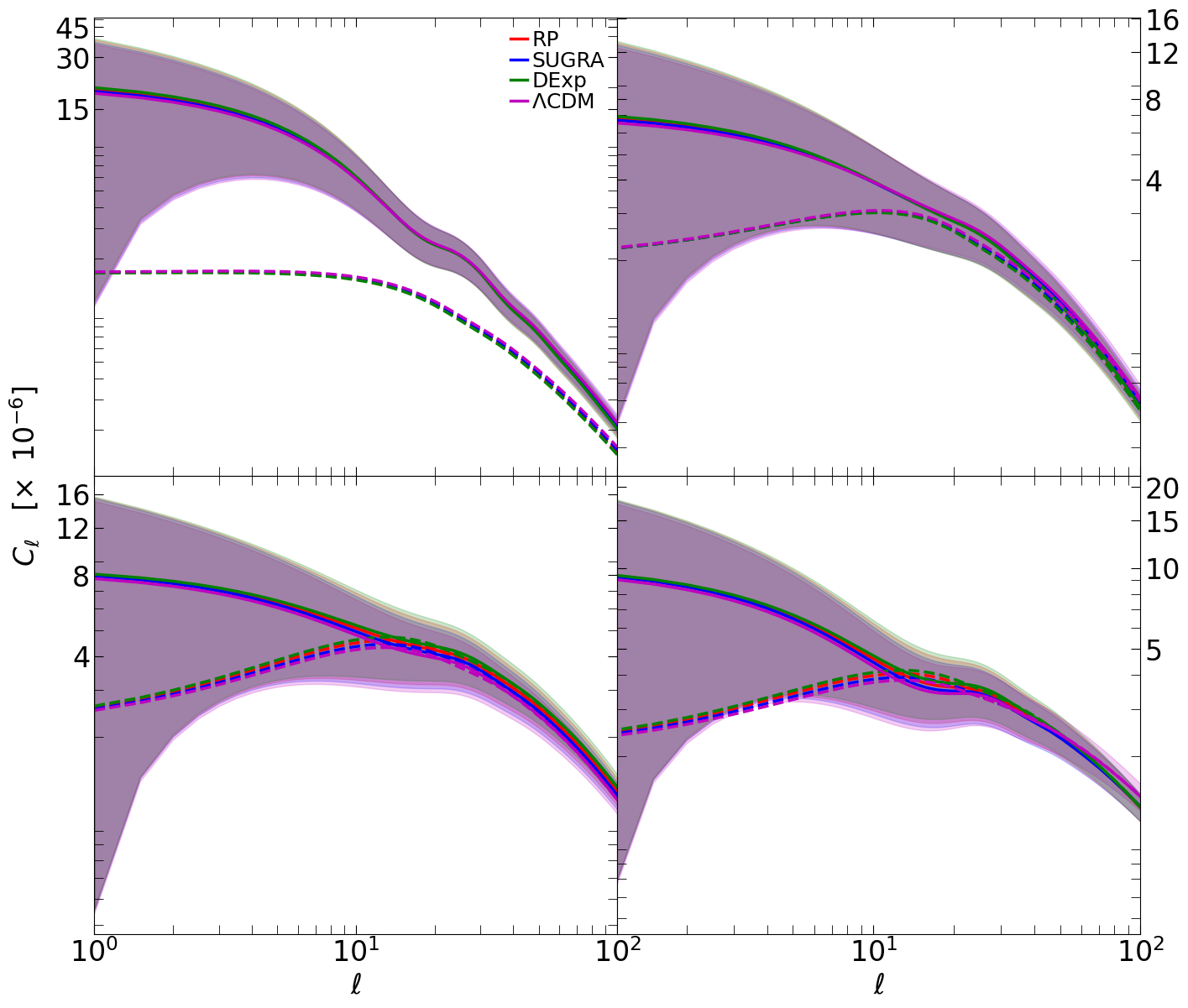}
\caption{The plots of the total magnification angular power spectrum $C_\ell$ (solid lines) and the lensing magnification angular power spectrum $C^{\rm lensing}_\ell$ (dashed lines), where $C^{\rm lensing}_\ell$ is computed with only the first integral in \eqref{f_ell} being considered: for RP \eqref{w_RP}, SUGRA \eqref{w_SUGRA}, and DExp \eqref{w_DExp} models, respectively, and that of the $\Lambda$CDM, at source redshifts $z_S \,{=}\, 0.5$ (top left), $z_S \,{=}\, 1$ (top right), $z_S \,{=}\, 3$ (bottom left), and $z_S \,{=}\, 4.5$ (bottom right). Shaded regions denote cosmic variance, on $C_\ell$.}\label{fig:totalCls}
\end{figure*}

In Fig.~\ref{fig:w_x}, we show the cosmic evolution of the equation of state parameters \eqref{w_RP}, \eqref{w_SUGRA}, and \eqref{w_DExp} for RP, SUGRA, and DExp quintessence models, respectively. We see that at redshifts $z \,{\lesssim}\, 1$ ($a \,{\gtrsim}\, 0.5$) the DExp model gives the highest equation of state parameter, followed by the RP and the SUGRA models, respectively. These results show the relative strengths of quintessence in the background cosmology: the farther away the equation of state parameter is from the cosmological constant solution $w_\Lambda \,{=}\, {-}1$, the stronger the evolution of quintessence, and hence the higher the contribution of quintessence in the total background energy density; consequently, the lower the amount of matter in the background energy density (given the conservation of total background energy). However, at higher redshifts $z \,{>}\, 1$ ($a \,{<}\, 0.5$), the quintessence equation of state parameters gradually converge to that of the cosmological constant, and hence evolve identically at the given regimes. We obtained the value of the equation of state parameters at today as $w_{\varphi 0} \simeq -0.848$ for RP, $w_{\varphi 0} \simeq -0.955$ for SUGRA, and $w_{\varphi 0} \simeq -0.819$ for DExp.


\subsection{The Magnification $C_\ell$'s}\label{subsec:Cls}
Here we examine the total magnification angular power spectrum \eqref{Cls}, and the lensing magnification angular power spectrum---noting lensing as the ``standard'' source of cosmic magnification. For all numerical computations we chose a logarithmic slope value $\alpha \,{=}\, {-}1.6$ so that we have a magnification bias value ${\cal Q} \,{=}\, 5$.

\begin{figure*}\centering
\includegraphics[scale=0.4]{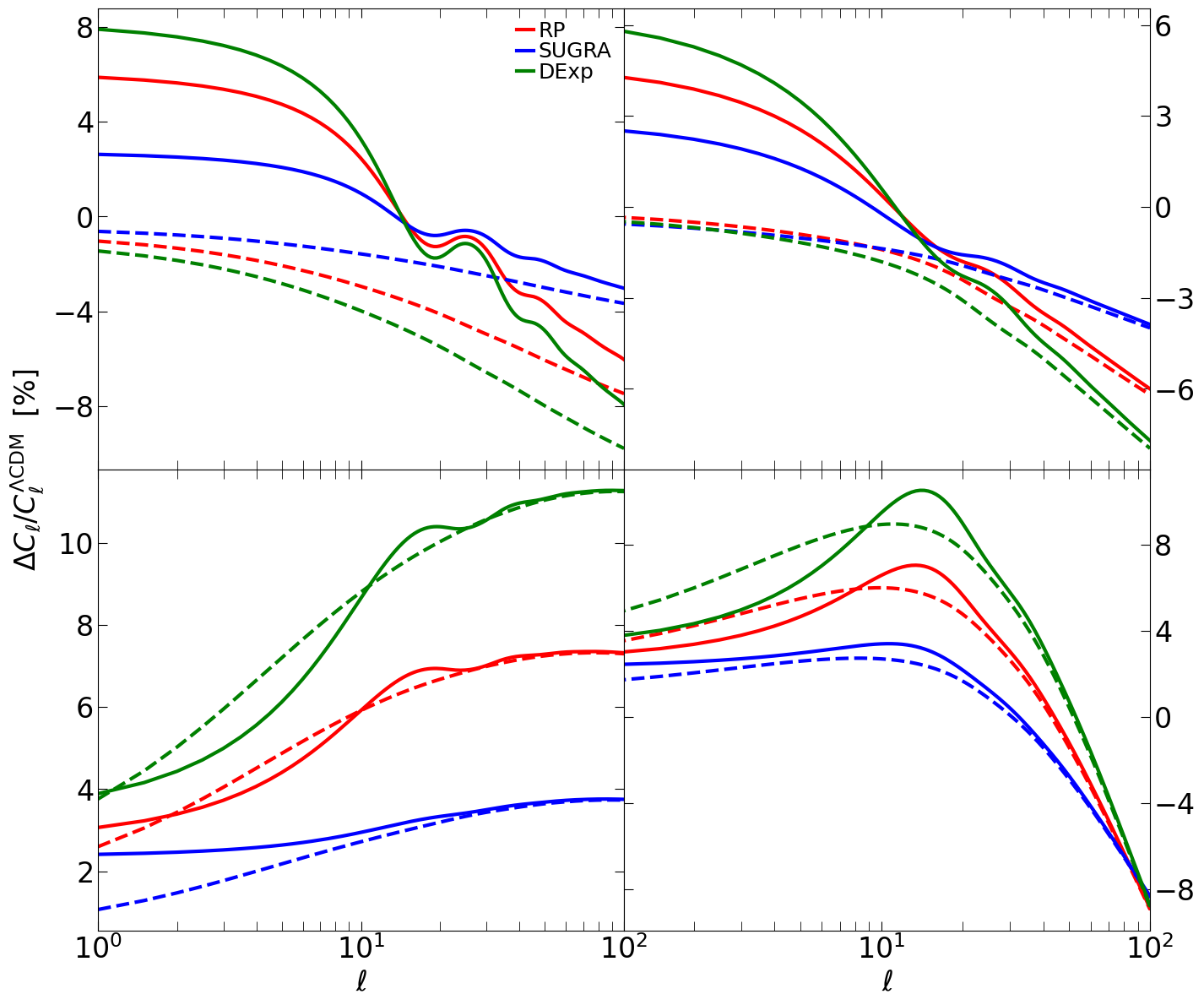}
\caption{The plots of fractional difference (in percentage) of the magnification angular power spectra of $\varphi$CDM relative to those of $\Lambda$CDM, at source redshifts: $z_S \,{=}\, 0.5$ (top left), $z_S \,{=}\, 1$ (top right), $z_S \,{=}\, 3$ (bottom left), and $z_S \,{=}\, 4.5$ (bottom right). Solid lines denote fractional difference of $C_\ell$, and dashed lines denote fractional difference of $C^{\rm lensing}_\ell$. Notations are as in FIG.~\ref{fig:totalCls}.}\label{fig:QCDM2LCDMpercentage}
\end{figure*}

In FIG.~\ref{fig:totalCls} we show the total magnification angular power spectrum $C_\ell$ (solid lines) and the lensing magnification angular power spectrum $C^{\rm lensing}_\ell$ (dashed lines), where $C^{\rm lensing}_\ell$ is computed with only the first integral in \eqref{f_ell} being considered for the three quintessence models \eqref{RP}, \eqref{SUGRA}, and \eqref{DExp}, with respect to multipole $\ell$, at source redshifts: $z_S \,{=}\, 0.5$ (top left panel), $z_S \,{=}\, 1$ (top right panel), $z_S \,{=}\, 3$ (bottom left panel), and $z_S \,{=}\, 4.5$ (bottom right panel). We also show the magnification angular power spectra of the well-known cosmological constant model ($\Lambda$CDM), at each $z_S$. We have also given the associated cosmic variance (shaded regions) for the total magnification angular power spectrum, at each $z_S$. We see that, in general, cosmic variance dominates over all the magnification angular power spectra of all the models, on all scales. In particular, cosmic variance grows on larger scales (smaller $\ell$). This implies that a lot of work will need to be done to isolate the total magnification signal in the angular power spectrum on the largest scales in the quantitaive analysis. Moreover, we observe that the angular power spectra ($C_\ell$ and $C^{\rm lensing}_\ell$) for the different models appear to be indistinguishable from one another on all scales, at the given $z_S$. This can be understandable, since the models have identical background evolutions for most of the cosmic history, until in the late-time universe. 

Moreover, all the models have the same physical sound speed $c_{s\varphi} \,{=}\, 1$, with $c_{s\varphi}$ prescribing the perturbation dynamics. Furthermore, we see that at $z_S \,{=}\, 3$ and $z_S \,{=}\, 4.5$ the lensing magnification angular power spectrum coincides with the total magnification angular power spectrum on scales $\ell \,{>}\, 10$, for all $\varphi$CDM models and for $\Lambda$CDM. However, the lensing magnification angular power spectrum deviates from with the total magnification angular power spectrum on scales $\ell \,{\lesssim}\, 10$ for all models, at all source redshifts. Thus, scales $\ell \,{\lesssim}\, 10$ will be suitable for investigating relativistic (non-lensing) effects in cosmic magnification. We also observe that, on going from $z_S \,{=}\, 4.5$ to $z_S \,{=}\, 0.5$, the lensing magnification angular power spectrum gradually exits the cosmic variance reach of the total magnification angular power spectrum. Thus, the lensing-magnification signal will not overlap with that of the total magnification at $z_S \,{\leq}\, 0.5$ in the observational data (or quantitative analysis). Conversely, at $z_S \,{>}\, 0.5$ (especially at $z_S \,{\geq}\, 1$), large-scale cosmic variations can cause the total-magnification signal to overlap with the lensing-magnification signal. Note that the lines (solid and dashed) in FIG.~\ref{fig:totalCls} give the mean values of the respective angular power spectra, while the shaded regions give the range of possible values of the amplitude of $C_\ell$ being governed by cosmic variance.

\begin{figure*}\centering
\includegraphics[scale=0.4]{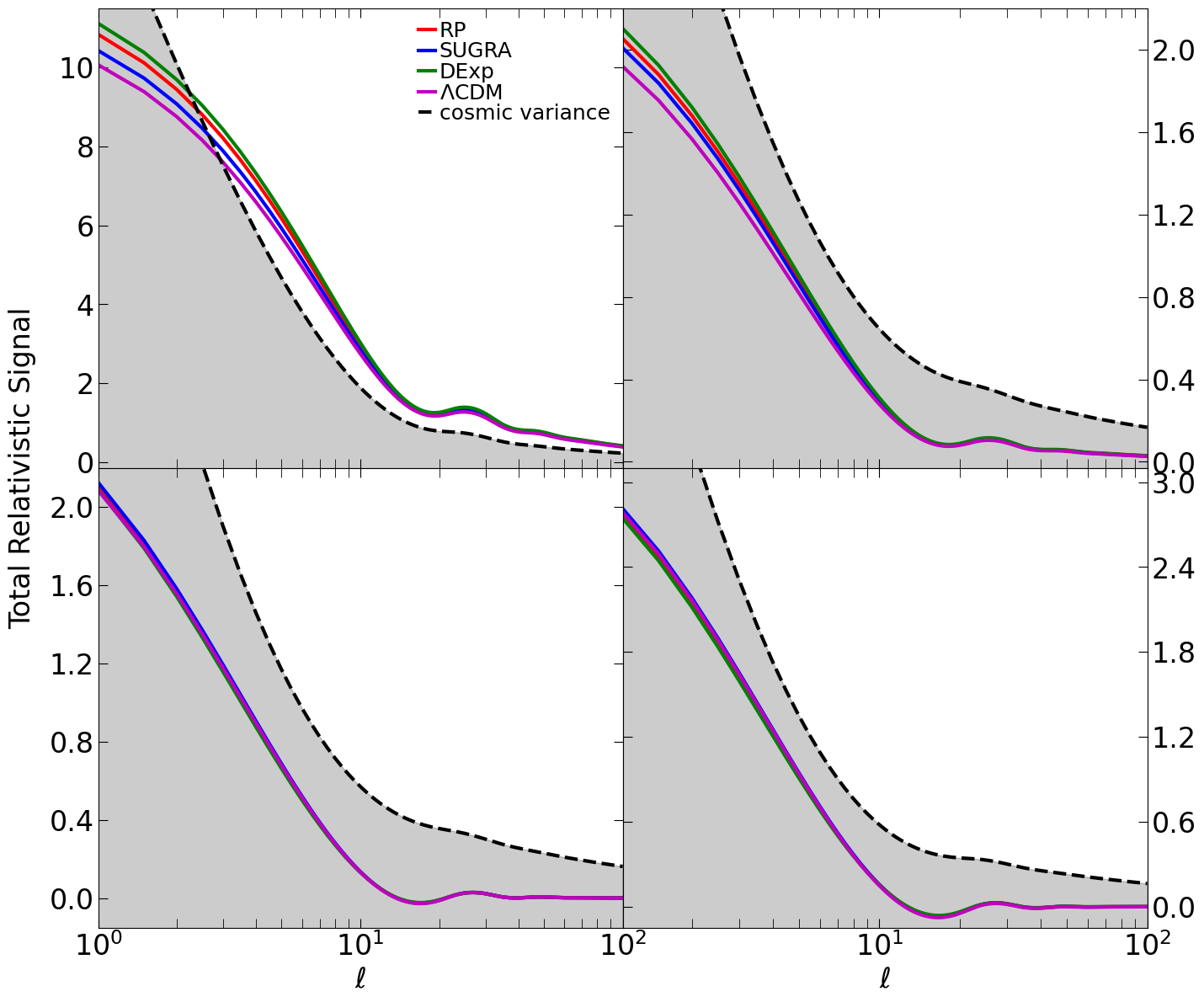}
\caption{The plots of the combined relativistic signal, $(C_\ell - C^{\rm lensing}_\ell) / C^{\rm lensing}_\ell$, in the total magnification angular power spectrum as a function of multipole $\ell$: at source redshifts $z_S \,{=}\, 0.5$ (top left), $z_S \,{=}\, 1$ (top right), $z_S \,{=}\, 3$ (bottom left), and $z_S \,{=}\, 4.5$ (bottom right). Shaded regions denote cosmic variance.}\label{fig:totalRelsSignal}
\end{figure*}

In FIG.~\ref{fig:QCDM2LCDMpercentage} we show the percentage fractional-difference between $\varphi$CDM and $\Lambda$CDM in the total magnification angular power spectrum $C_\ell$ (solid lines) and the lensing magnification angular power spectrum $C^{\rm lensing}_\ell$ (dashed lines), as functions of multipole $\ell$, at source redshifts: $z_S \,{=}\, 0.5,\, 1,\, 3,\, 4.5$. (Notations and style are as in FIG.~\ref{fig:totalCls}.) The plots compare the quintessence models to the (standard) cosmological constant model in the magnification angular power spectra. Since the quintessence models are compared to the same model ($\Lambda$CDM), the separation between the lines measures the difference between the models, and hence gives an indication of the possibility of distinguishing between the quintessence models. On the other hand, the percentage value measures the similarity (or dissimilarity) to $\Lambda$CDM. Thus, the closer the percentage is to zero, the more difficult it will be to distinguish the model from $\Lambda$CDM. We see that at $z_S \,{\leq}\, 1$, while the separation between the percentage differences of the total magnification angular power spectrum is more prominent and increasing on larger scales (smaller $\ell$), that of the lensing magnification angular power spectrum is smaller and converging. This shows that the total-magnification signal is more sensitive to the underlying dark energy model than the lensing-magnification signal. In other words, the magnification angular power spectrum with relativistic corrections \eqref{Doppler}--\eqref{potential} taken into account is more sensitive to parameters (and their changes) in the underlying dark energy model. This can be crucial for placing constraints, and possibly distinguishing quintessence models, in the quantitative or observational analysis, at the given redshifts. Conversely, it will be more difficult to distinguish quintessence with only lensing magnification, at $z_S \,{\leq}\, 1$. (See also analysis in e.g. \cite{Duniya:2016gcf, Duniya:2023xgx, Duniya:2022vdi, Duniya:2022xcz}.) Moreover, given that the amplitude of the lensing magnification angular power spectrum percentage differences approaces zero on the largest scales, it implies that it will be difficult to differentiate quintessence from the cosmological constant in the lensing magnification angular power spectra on these scales, at the given $z_S$.

\begin{figure*}\centering
\includegraphics[scale=0.4]{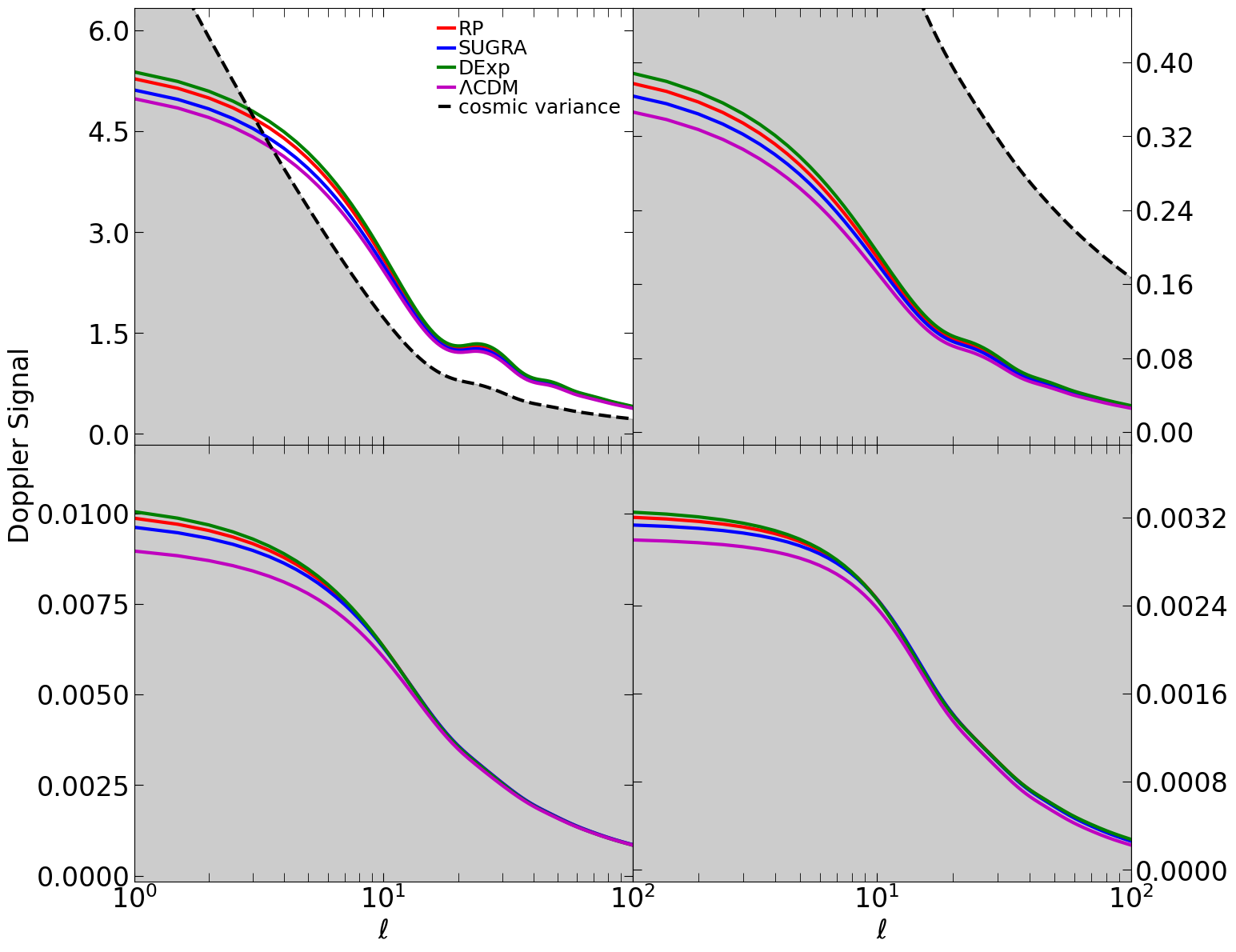}
\caption{The plots of the Doppler signal, $(C_\ell - C^{(\rm no\, Doppler)}_\ell) / C^{(\rm no\, Doppler)}_\ell$, in the total magnification angular power spectrum as a function of multipole $\ell$: at source redshifts $z_S \,{=}\, 0.5$ (top left), $z_S \,{=}\, 1$ (top right), $z_S \,{=}\, 3$ (bottom left), and $z_S \,{=}\, 4.5$ (bottom right). Shaded regions denote cosmic variance.}\label{fig:DopplerSignal}
\end{figure*}

Furthermore, at $z_S \,{\geq}\, 3$ we observe that for each quintessence model, the percentage difference of the lensing-magnification signal is (fairly) of the same order as those of the total-magnification signal, especially on scales $\ell \,{>}\, 20$. Similarly for the separation between the percentage difference. This is consistent with the results in FIG.~\ref{fig:totalCls}. Thus, at $z_S \,{\geq}\, 3$, either lensing magnification or total cosmic magnification can be used in comparing or distinguishing quintessence from the cosmological constant, and distinguishing quintessence models from one another. In other words, the lensing magnification angular power spectrum can be taken as a reasonable approximation of the total magnification angular power spectrum, at $z_S \,{>}\, 3$. On the other hand, at $z_S \,{<}\, 3$ (specifically, at $z_S \,{\leq}\, 1$) there is a reasonable separation of the percentage differences of the lensing-magnification signal from those of the total-magnification signal, with the percentage differences of the total-magnification signal gradually dominating over those of the lensing-magnification signal on the largest scales ($\ell \,{\leq}\, 10$): the percentage differences of the lensing-magnification signal are all negative; whereas, those of the total-magnification signal are all positive. This implies that $\Lambda$CDM predicts more lensing magnification than $\varphi$CDM, and conversely, $\varphi$CDM predicts more total cosmic magnification than $\Lambda$CDM, at the given epochs. It is also worth noting that the actual percentage values change at the different $z_S$ for both $C_\ell$ and $C^{\rm lensing}_\ell$. It is already known (see e.g. \cite{Duniya:2015nva}) that cross-correlation terms of relativistic corrections in the power spectrum can give alternating (positive or negative) contributions at low or high $z_S$; thereby changing the amplitude of the power spectrum. Moreover, the amplitude of lensing (as well as ISW effect and time delay), being an integral effect, will increase with an increase in $z_S$---as radial distance increases.

\begin{figure*}\centering
\includegraphics[scale=0.4]{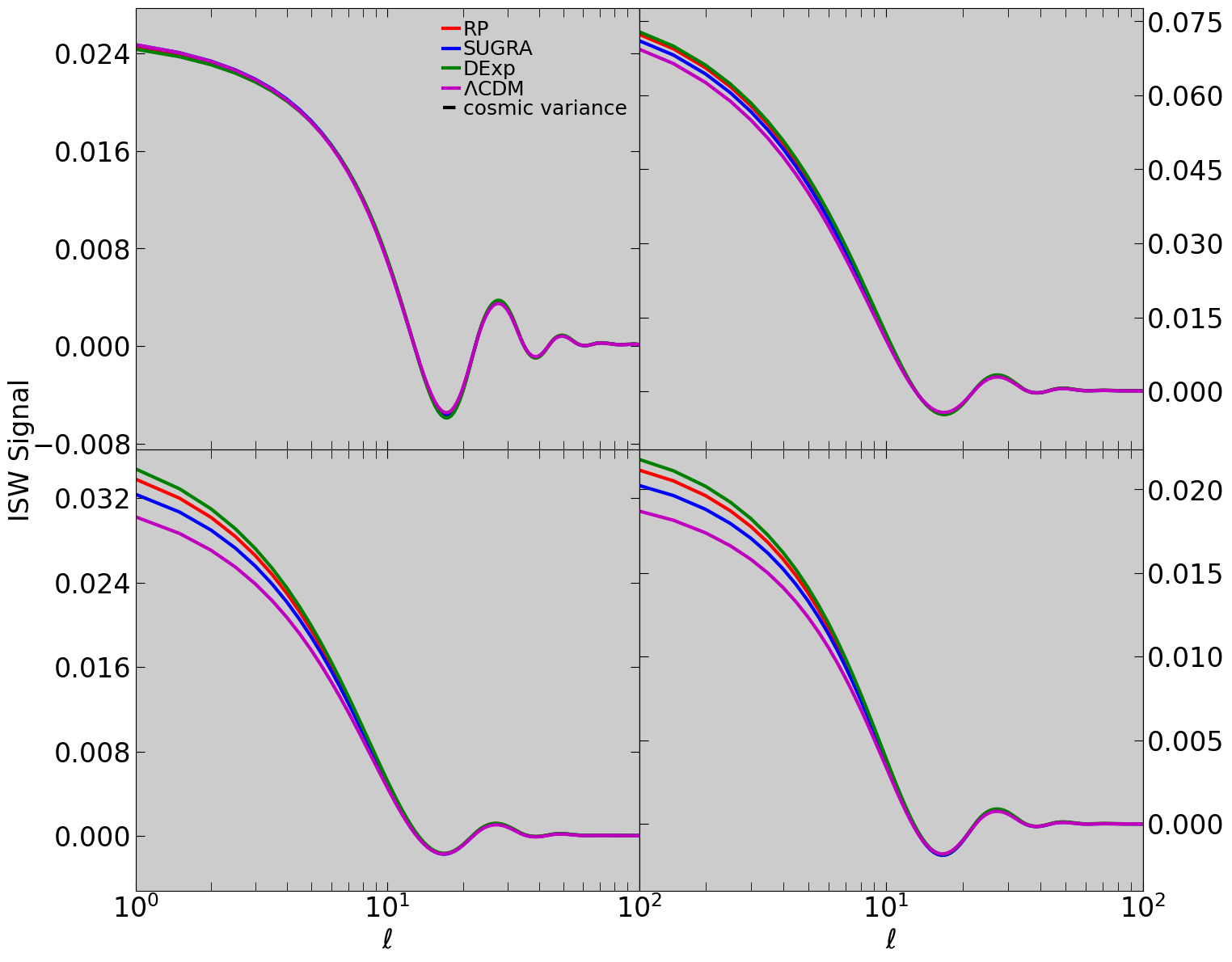}
\caption{The plots of the ISW signal, $(C_\ell - C^{(\rm no\, ISW)}_\ell) / C^{(\rm no\, ISW)}_\ell$, in the total magnification angular power spectrum as a function of multipole $\ell$: at source redshifts $z_S \,{=}\, 0.5$ (top left), $z_S \,{=}\, 1$ (top right), $z_S \,{=}\, 3$ (bottom left), and $z_S \,{=}\, 4.5$ (bottom right). Shaded regions denote cosmic variance.}\label{fig:ISWSignal}
\end{figure*}

In FIG.~\ref{fig:totalRelsSignal} we show plots of the fractional change, $C_\ell / C^{\rm lensing}_\ell - 1$, in the total magnification angular power spectrum owing solely to relativistic corrections \eqref{Doppler}--\eqref{potential}, with respect to multipole $\ell$: at source redshifts $z_S \,{=}\, 0.5,\, 1,\, 3,\, 4.5$. (Notations and style are as in FIG.~\ref{fig:totalCls}.) These fractions measure the total relativistic signal in the total magnification angular power spectrum for $\varphi$CDM and $\Lambda$CDM, at the given source redshifts. By going from high to low redshifts, i.e. from $z_S \,{=}\, 4.5$ (bottom right) to $z_S \,{=}\, 0.5$ (top left), we observe that the amplitude of the total relativistic magnification signal for each quintessence model, and the cosmological constant model, increases. This suggests that the amount of total relativistic signal in cosmic magnification will be higher at low source redshifts ($z_S \,{\leq}\, 0.5$) than at high source redshifts ($z_S \,{>}\, 0.5$) in both $\varphi$CDM and $\Lambda$CDM. This is consistent with results in FIGS.~\ref{fig:totalCls} and~\ref{fig:QCDM2LCDMpercentage} (see also \cite{Duniya:2016gcf, Duniya:2023xgx}). Moreover, we see that the total relativistic magnification signal for all models gradually becomes significant with respect to cosmic variance as $z_S$ decreases: at $z_S \,{\leq}\, 1$ the amplitude is below cosmic variance on all scales, and only surpasses cosmic variance at $z_S \,{=}\, 0.5$. This suggests that the effect may be detectable at the given $z_S$ (subject proper measurement error analysis; not considered in this work). In other words, neglecting cosmic variance in the quantitative analysis of the total relativistic signal in the magnification angular power spectrum will not lead to a significant deviation in estimations, at $z_S \,{\leq}\, 0.5$. However, at $z_S \,{>}\, 0.5$, advanced analytical methods like multi-tracer methods (e.g.~\cite{Abramo:2013awa}) will need to be included in the analysis to beat down cosmic variance, in order to stand a chance to isolate the total relativistic signal in the magnification angular power spectrum. Furthermore, the fact that the plots are closely spaced on all scales suggests that it will be difficult to distinguish dark energy models with the total relativistic signal in the total magnification angular power spectrum, at all source redshifts. 

\begin{figure*}\centering
\includegraphics[scale=0.4]{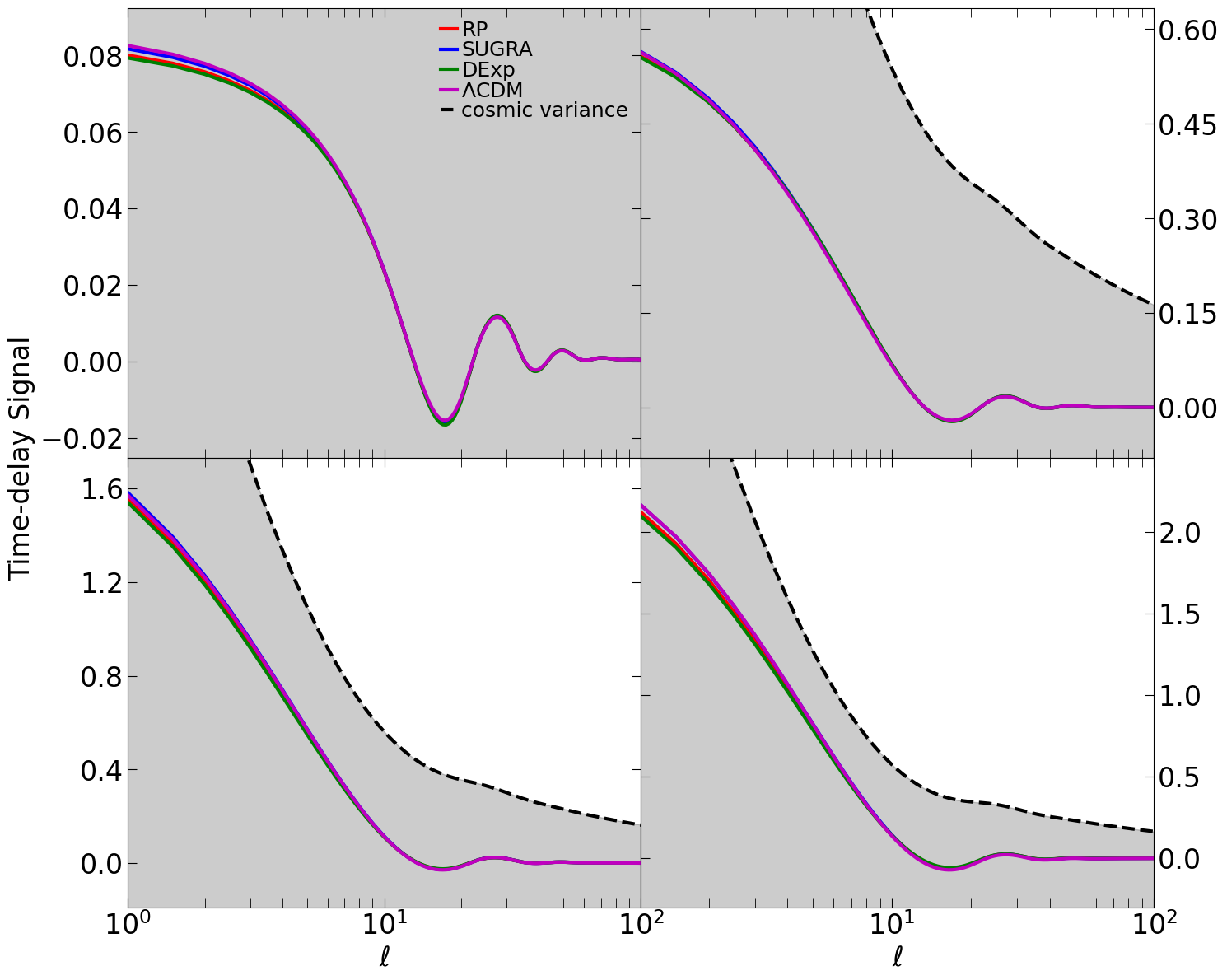}
\caption{The plots of the time-delay signal, $(C_\ell - C^{(\rm no\, timedelay)}_\ell) / C^{(\rm no\, timedelay)}_\ell$, in the total magnification angular power spectrum as a function of multipole $\ell$: at source redshifts $z_S \,{=}\, 0.5$ (top left), $z_S \,{=}\, 1$ (top right), $z_S \,{=}\, 3$ (bottom left), and $z_S \,{=}\, 4.5$ (bottom right). Shaded regions denote cosmic variance.}\label{fig:timedelaySignal}
\end{figure*}

Similarly, in FIG.~\ref{fig:DopplerSignal} we show plots of the fractional change, $C_\ell / C^{(\rm no\, Doppler)}_\ell - 1$, in the total magnification angular power spectrum owing solely to Doppler-effect correction \eqref{Doppler}, with respect to multipole $\ell$: at source redshifts $z_S \,{=}\, 0.5,\, 1,\, 3,\, 4.5$. (Notations and style are as in FIG.~\ref{fig:totalCls}.) These fractions measure the Doppler signal in the total magnification angular power spectrum for $\varphi$CDM and $\Lambda$CDM, at the given source redshifts. We observe that the Doppler magnification signal is almost identical to the total relativistic magnification signal, except that the amplitude of the Doppler magnification signal is relatively lower, for each quintessence model, and the cosmological constant model. Thus, similar discussion follows. The results suggest that the Doppler signal dominates in the total relativistic magnification signal at low source redshifts ($z_S \,{\leq}\, 0.5$), in both $\varphi$CDM and $\Lambda$CDM. This is consistent with results in e.g. \cite{Duniya:2016gcf, Duniya:2023xgx}. 

In FIG.~\ref{fig:ISWSignal} we show plots of the ISW signal, $C_\ell / C^{(\rm no\, ISW)}_\ell - 1$, in the total magnification angular power spectrum, with respect to multipole $\ell$: at source redshifts $z_S \,{=}\, 0.5,\, 1,\, 3,\, 4.5$. We give the ISW magnification signal for $\varphi$CDM and $\Lambda$CDM, at the given source redshifts. We observe that the signal is below cosmic variance on all scales, at all source redshifts. This is contrary to the results e.g. for unified dark energy \cite{Duniya:2023xgx}, where it was found that the ISW magnification signal becomes significant relative to cosmic variance, at $z_S \,{\geq}\, 3$. Here (FIG.~\ref{fig:ISWSignal}), although the ISW magnification signals for the different dark energy models begin to differentiate on scales $\ell \,{<}\, 10$ at $z_S \,{\geq}\, 3$, the amplitudes still remain below cosmic variance. Thus, without advanced methods (including multi-tracer analysis), it will be difficult to extract the ISW magnification signal for quintessence, and the cosmological constant, at all source redshifts. 

Similarly, in FIG.~\ref{fig:timedelaySignal} we show plots of the time-delay magnification signal, $C_\ell / C^{(\rm no\, timedelay)}_\ell - 1$, with respect to multipole $\ell$: at source redshifts $z_S \,{=}\, 0.5,\, 1,\, 3,\, 4.5$. We see similar behaviour in the time-delay magnification signal as in the the ISW magnification signal (FIG.~\ref{fig:ISWSignal}) for $\varphi$CDM and $\Lambda$CDM, at the given source redshifts. The only difference is in the amplitudes: the time-delay magnification signals are larger in amplitude relative to the ISW magnification signals, at all source redshifts. Nevertheless, the time-delay magnification signals are also below cosmic variance for all the dark energy models, at all the given source redshifts. Thus, any quantitative analysis of time-delay signal in the total magnification angular power spectrum will also require advanced methods like multi-tracer analysis.

For the gravitational-potential correction \eqref{potential}, the magnification signal is zero on all scales, at all the given source redshifts (as in FIGS.~\ref{fig:totalCls}--\ref{fig:timedelaySignal}). Thus, we do not show those results here.


\section{Conclusion}\label{sec:Concl}
We presented a comprehensive analysis of cosmic magnification in a universe dominated by quintessence and cold dark matter ($\varphi$CDM), using three well-known quintessence models. We used the angular power spectrum as the statistic of choice. We discussed the total magnification angular power spectrum and the (standard) lensing magnification angular power spectrum for the three quintessence models, at several source redshifts $z_S$. Moreover, we compared the quintessence models to the cosmological constant model ($\Lambda$CDM) in both the total magnification angular power spectrum and the lensing magnification angular power spectrum. Furthermore, we examine the relativistic (non-lensing) signals, combined and individually, in the magnification angular power spectrum.

We found that it will be difficult to isolate the total magnification signal in the angular power spectrum in the quantitaive analysis. We also found that quintessence will be undistiguishable from the cosmological constant in both the total magnification angular power spectrum and the lensing magnification angular power spectrum on all scales, at high source redshifts ($z_S \,{\geq}\, 3$). The results suggest that scales $\ell \,{\lesssim}\, 10$ will be suitable for analysing relativistic effects in cosmic magnification, at $z_S \,{\leq}\, 0.5$, as relativistic magnification signals are relatively easily differentiable from lensing-magnification signal. On the other hand, at $z_S \,{>}\, 0.5$, large-scale cosmic variations can cause the total-magnification signal to overlap with the lensing-magnification signal, and therefore making it difficult to disentagle lensing-magnification signal from relativistic magnification signals. 

Moreover, we found that at $z_S \,{\geq}\, 3$, lensing magnification angular power spectrum can suitably approximate the total cosmic magnification angular magnification power spectrum. Our results also suggest that $\varphi$CDM will predict less lensing magnification than $\Lambda$CDM, but will predict more total cosmic magnification than $\Lambda$CDM, at $z_S \,{\leq}\, 1$. Moreover, at $z_S \,{\leq}\, 1$, the magnification angular power spectrum with relativistic corrections taken into account showed more sensitivity to the quintessence parameters. This can be crucial for placing constraints, and possibly distinguishing dark energy models, in the data, at the given source redshifts. 

Furthermore, we found that the total relativistic magnification signal surpass cosmic variance at low source redshifts ($z_S \,{\leq}\, 0.5$) in both $\varphi$CDM and $\Lambda$CDM: consistent with results in e.g. \cite{Duniya:2016gcf, Duniya:2023xgx}. Thus, this effect may be detectable at the given $z_S$---subject to proper measurement error analysis. Alternatively, it implies that neglecting cosmic variance in the analysis of the total relativistic magnification signal will not be of significant consequence, at the given $z_S$. However, at $z_S \,{>}\, 0.5$ cosmic variance cannot be neglected, and advanced analytical methods including multi-tracer methods will required in the analysis to beat down cosmic variance---in order to increase the possibility of isolating the total relativistic signal in the magnification angular power spectrum. Similar discussion follows for the Doppler magnification signal.

The ISW and the time-delay magnification signals, respectively, appeared to be surpassed by cosmic variance for both quintessence and the cosmological constant on all scales, at the source redshifts ($z_S \,{\leq}\, 4.5$). Thus, without multi-tracer analysis it will be difficult to extract the ISW and the time-delay magnification signals in $\varphi$CDM and $\Lambda$CDM. The gravitational-potential magnification signal in both $\varphi$CDM and $\Lambda$CDM becomes zero, at the same source redshifts ($z_S \,{\leq}\, 4.5$).

\begin{acknowledgments}
We thank Roy Maartens for useful comments. We also thank the Centre for High Performance Computing, Cape Town, South Africa, for providing the computing facilities with which all the numerical computations in this work were done. EM has received funding from the Pan-African Planetary and Space science Network (PAPSSN). PAPSSN is founded by the Intra-Africa Academic Mobility Scheme of the European Union under the grant agreement No. 624224.\\
\end{acknowledgments}

\section*{DATA AVAILABILITY}
	Data sharing is not applicable to this article, as no data were created or analyzed.

\appendix

\section{Gauge Invariance }
\label{App:Inv}
\subsection{Gauge-invariant 4-velocity}
 Here we show that the gauge-invariant perturbation of the 4-velocity, $\delta \hat{u}^\beta$, is constructed by 
\begin{align}
	\delta\hat{u}^\beta &= \delta {u}^\beta - \L_{\xi}\bar{u}^\beta,\\ \nonumber
	&= \delta {u}^\beta - \xi^\nu \partial_\nu \bar{u}^\beta +\bar{u}^\nu \partial_{\nu}\xi^\beta,\\ \nonumber
	&= \delta {u}^\beta - \xi^0 (\bar{u}^0)' \delta^{\beta}{}_{0}+ \bar{u}^0 (\xi^0 \delta^{\beta}{}_{0}+\xi^i \delta^\beta{}_{i})',\\ \nonumber
	&= \left[ \delta u^0 - \xi^0 (\bar{u}^0)' + \bar{u}^0 (\xi^0)' \right] \delta^\beta{}_{0}
	+ \left[ \delta u^i + (\xi^i)' \bar{u}^0 \right] \delta^\beta{}_{i},\\ \nonumber
	&= \left[ -a^{-1} \phi - \xi^0 (a^{-1})' + a^{-1} \xi^{0\prime} \right] \delta^\beta{}_{0}\\ \nonumber
	&+ \left[ a^{-1} v^i + a^{-1} (\xi^{i})^\prime \right] \delta^\beta{}_{i}, \nonumber \\
	&= -a^{-1} \left( \phi - \mathcal{H} \xi^0 - \xi^{0\prime} \right) \delta^\beta{}_{0}
	+ a^{-1} \left( \partial^i v + \partial^i \xi^{\prime} \right) \delta^\beta{}_{i}, \nonumber \\
	&= a^{-1} \left( -\phi + \mathcal{H}\xi^0 + \xi^{0\prime}, \; \partial^i \left( v + \xi^{\prime} \right) \right), \nonumber \\
	&= a^{-1} \left( -\hat{\phi}, \; \partial^i \hat{v} \right),
\end{align}
 where it is easy to show $\delta u^0 = -a^{-1}\phi$, with $\delta u^i = a^{-1}v^i$ (Note that $v^i$ is taken as a purely scalar-generated vector, i.e. $v^i = \partial^i v$.) Thus, we have 
\begin{align}
	\hat{u}^\beta = \bar{u}^\beta + \delta \hat{u}^\beta = a^{-1}\delta^\beta{}_{0}+\delta \hat{u}^\beta,
\end{align}
 and we have
\begin{align}
	\hat{u}^\beta= a^{-1}(1-\hat{\phi},\partial^i \hat{v}).
\end{align}
 Then, by applying (\ref{gauge2}),
\begin{align}
	\hat{u}^\beta \rightarrow {U}^\beta = a^{-1}(1-\Phi,\partial^i V),
\end{align}
 where $\Phi$ and $V$ are as in (\ref{inv-Phi}) and (\ref{inv-V}).
\subsection{Gauge-invariance of Phi, Psi, and V}
To show that the potentials $\Phi$, $\Psi$, and $V$, given in \eqref{inv-Phi}--\eqref{inv-V}, are gauge-invariant we only need to show that $\hat{\Phi} = \Phi$ (similarly for $\Psi$, and $V$).

For $\Phi$, we have
\bea\label{invPhi}
\hat{\Phi} &=& \hat{\phi} - {\cal H}(\hat{E}'-\hat{B}) + \hat{B}' - \hat{E}'' \nn
&=& \phi - {\cal H}\xi^0 - \xi^0{'} - {\cal H} (E' - \xi' - B - \xi^0 + \xi') \nn
&& +\; B' + \xi^0{'} - \xi'' - E'' + \xi'' \nn
&=& \phi -{\cal H}(E'-B) + B' -E'' \nn
&=& \Phi.
\eea

For $\Psi$, we have
\bea\label{invPsi}
\hat{\Psi} &=& \hat{D} +\dfrac{1}{3}\nabla^2\hat{E} + {\cal H}(\hat{E}'-\hat{B}) \nn
&=& D +\dfrac{1}{3}\nabla^2\xi + {\cal H}\xi^0 + \dfrac{1}{3}\nabla^2 (E-\xi) \nn
&& +\; {\cal H}(E'-\xi' - B - \xi^0 + \xi') \nn
&=& D + \dfrac{1}{3}\nabla^2E + {\cal H}(E'-B) \nn
&=& \Psi.
\eea

For $V$, we have
\bea\label{invV}
\hat{V} &=& \hat{v} + \hat{E}' \nn
&=& v+\xi' + E' -\xi' \nn
&=& V.
\eea
Thus, the potential $\Phi$, $\Psi$, and $V$ as given in \eqref{inv-Phi}--\eqref{inv-V} are in fact gauge-invariant.
\subsection{Redshift-space area density contrast}
 Here we show that the redshift-space area density contrast $\delta\tilde{\cal A}/\bar{\cal A}$ is gauge-invariant, using \eqref{delta-A} and \eqref{delta-z}: 
\begin{align}\label{deltaA}
	\dfrac{\delta\tilde{\cal A}}{\bar{\cal A}} =&\, \dfrac{\delta{\cal A}}{\bar{\cal A}} + 2\Big(1-\dfrac{1}{{\cal H}\bar{r}}\Big) \dfrac{\delta{z}}{1+\bar{z}}, \\
	=& \int^{\bar{r}_S}_0 d\bar{r} \dfrac{\bar{r} - \bar{r}_S}{\bar{r}_S \bar{r}} \nabla^2_\Omega \Big(\phi + D + \dfrac{1}{3}\nabla^2 E + B' - E''\Big) \nn
	+& \dfrac{2}{\bar{r}_S} \int^{\bar{r}_S}_0 d\bar{r} \Big(\phi + D + \dfrac{1}{3}\nabla^2 E + B' - E''\Big) \nn
	+& \Big[\dfrac{2}{\bar{r}_S } \left(B-E'\right) -2D - \dfrac{2}{3}\nabla^2 E \Big]^S_O  \nn
	+& 2\Big(1 {-} \dfrac{1}{{\cal H}\bar{r}}\Big) \int^{\bar{r}_S}_0 d\bar{r} \Big[\phi' + D' + \dfrac{\nabla^2}{3} E' + (B' {-} E'')'\Big], \nn
	-& 2\Big(1 {-} \dfrac{1}{{\cal H}\bar{r}}\Big) \Big[\phi + B' - E'' - \dfrac{\partial V}{\partial\bar{r}} \Big]^S_O , \\
	=& {-}2\Psi\Big|^S_O + \int^{\bar{r}_S}_0{ d\bar{r}\left[\dfrac{2}{\bar{r}_S} + \Big(\bar{r} - \bar{r}_S\Big) \dfrac{\bar{r}}{\bar{r}_S } \nabla^2_\perp \right] \left(\Phi + \Psi\right) } \nn
	-& 2\Big(1 {-} \dfrac{1}{\bar{r}_S {\cal H}}\Big) \Big[\Phi + \Psi - D - \dfrac{\nabla^2}{3} E + {\cal H}(B-E')\Big]^S_O \nn
	+& 2\Big(1 {-} \dfrac{1}{\bar{r}_S {\cal H}}\Big) \Big[ \dfrac{\partial{V}}{\partial\bar{r}} \Big|^S_O + \int^{\bar{r}_S}_0 d\bar{r} \left(\Phi' +\Psi'\right)\Big],\\
	=& {-}2\Psi\Big|^S_O + \int^{\bar{r}_S}_0{ d\bar{r}\left[\dfrac{2}{\bar{r}_S} + \Big(\bar{r} - \bar{r}_S\Big) \dfrac{\bar{r}}{\bar{r}_S } \nabla^2_\perp \right] \left(\Phi + \Psi\right) } \nn
	+& 2\Big(1 {-} \dfrac{1}{\bar{r}_S {\cal H}}\Big) \Big[ \dfrac{\partial{V}}{\partial\bar{r}} \Big|^S_O+\int^{\bar{r}_S}_0 d\bar{r} \left(\Phi' +\Psi'\right)\Big]\nn
	-& 2\Big(1 {-} \dfrac{1}{\bar{r}_S {\cal H}}\Big) \Phi\Big|^S_O  ,
\end{align}
 where $\nabla^2_\Omega = \bar{r}^2 \nabla^2_\perp$, and $\bar{\cal A}$, $\bar{r}_S$, $\nabla^2_\perp$, $\Phi$, $\Psi$, and $V$ are as in \S\ref{sec:overdensity}.

Alternatively, the gauge-invariance of the redshift-space area density (\ref{deltaA}) can also be shown generically as follows:

From (\ref{mu}) and (\ref{dA-trans}). To show that $\delta\tilde{\cal A}/\bar{\cal A}$ is gauge-invariant, we only need to show that $\delta\hat{\cal\tilde A}/\bar{\cal A} = \delta\tilde{\cal A}/\bar{\cal A}$, therefore  
\begin{align}
	\frac{\delta \hat{ \tilde{\mathcal{A}}}}{\bar{\mathcal{A}}} &= \frac{\delta \hat{{\mathcal{A}}}}{\bar{\mathcal{A}}}+2 \left( 1 - \frac{1}{\mathcal{H} \bar{r}} \right) \frac{\delta  \hat{z}}{1 + \bar{z}},\nn
	&= \frac{\delta \hat{{\mathcal{A}}}}{\bar{\mathcal{A}}}+ \frac{d \ln \bar{\mathcal{A}}}{d  \bar{z}}{\delta \hat{z}},\nn
	&= \frac{\delta {\mathcal{A}}}{\bar{\mathcal{  A}}} - \xi^{0}\frac{\bar{\mathcal{A}}'}{\bar{\mathcal{A}}}+ \frac{d \ln \bar {\mathcal{A}}}{d \bar{z}}\left[ \delta z - \xi^0 \bar{z}'  \right],\nn
	&= \frac{\delta \mathcal{A}}{\bar{\mathcal{A}}} + \frac{d \ln\mathcal{A}}{d\bar{z}} \delta z,\nn
	&= \frac{\delta \tilde{\mathcal{A}}}{\bar{\mathcal{A}}},
\end{align}
where we used (\ref{dA-trans}), and (\ref{eq:A/z}),  and 
\begin{align}
	\frac{\bar{\mathcal{A}}'}{\bar{\mathcal{A}}} &= \frac{\partial \bar{z}}{\partial \bar{\eta}} \frac{d \ln \bar{\mathcal{A}}}{d \bar{z}} = \bar{z}'  \frac{d \ln \bar{\mathcal{A}}}{d \bar{z}}.
\end{align}
This proves that the redshift-space area density contrast $\delta \tilde{\mathcal{A}}/\bar{\mathcal{A}}$ is gauge-invariant. It can also be shown by the invariance  of $\Phi$, $\Psi$, and $V$, as given below.

\section{The Cosmological Equations}\label{QCDM:Eqns}

In a late-time universe dominated by matter (dark plus baryonic) and quintessence $\varphi$, which is driven by a potential $U(\varphi)$, the Friedmann equation is given by
\bea\label{Friedmann}
\mathcal{H}^2 &=& \dfrac{8{\pi}Ga^{2}}{3}\left[\rho_m +\dfrac{{\varphi}'^2}{2a^2} + U(\varphi)\right], \nonumber\\
&=& \left(\Omega_m + \Omega_\varphi\right) {\cal H}^2, 
\eea
where ${\cal H}$ is as in \eqref{inv-Phi}, and the Raychaudhuri equation, is
\beq\label{Raychaudhuri}
{\cal H}' = -\dfrac{1}{2} \left(1+3w_\varphi\Omega_\varphi\right) {\cal H}^2,
\eeq
where $w_\varphi$ is as in \eqref{w_q}, with the evolution of quintessence being given by \eqref{d2Vardt2}.

The gravitational potential evolves according to
\beq\label{dPhidt}
\Phi' +{\cal H}\Phi = -{3\over 2}{\cal H}^2\left[\Omega_m V_m + \Omega_\varphi (1+w_\varphi) V_\varphi\right],
\eeq
where the 4-vectors are
\beq
U^\beta_m = a^{-1}(1-\Phi,\partial^i V_m),\quad U^\beta_\varphi = a^{-1}(1-\Phi,\partial^i V_\varphi),
\eeq
with all definitions as in \S\ref{sec:overdensity}.

The matter fluctuations obey energy-momentum conservation, given by
\bea\label{dDmdt}
\Delta'_{m} - {9\over 2}{\cal H}^{2}\Omega_\varphi (1+w_\varphi)\left( V_m-V_\varphi\right) &=& -\nabla^{2}V_m, \\
\label{dVmdt}
V'_m + {\cal H}V_m &=& -\Phi,
\eea
and for quintessence, we have
\bea\label{dDxdt}
\Delta'_\varphi - 3w_\varphi{\cal H}\Delta_\varphi &-& \dfrac{9}{2}{\cal H}^2 \Omega_m(1+w_\varphi)\left(V_\varphi-V_m\right)  \nonumber\\
 &=& -(1+w_\varphi)\nabla^2V_\varphi,\\ \label{dVxdt}
 V'_\varphi + {\cal H}V_\varphi &=& -\Phi - \dfrac{c^2_{s\varphi}}{1+w_\varphi}\Delta_\varphi ,
\eea
where $c_{s\varphi}$ is the dark energy sound speed. Note that \eqref{dPhidt}--\eqref{dVxdt} hold for any form of dark energy, with one simply having to specify $w_\varphi$ and $c_{s\varphi}$ (See \S\ref{sec:QCDM}.

\bibliography{magnification_in_quintessence}

\end{document}